\documentclass[aps,prl,reprint,showpacs,footinbib,amsmath,amssymb,amsfonts,floatfix,superscriptaddress]{
revtex4-1}
\usepackage{graphicx}
\usepackage{epsfig}
\usepackage{hyperref}

\begin{document}

\author{A. K. Karlis}
\email{karlis@physi.uni-heidelberg.de}
\affiliation{Department of Physics, University of Athens, GR-15771 Athens, Greece}
\author{F. K. Diakonos}
\email{fdiakono@phys.uoa.gr}
\affiliation{Department of Physics, University of Athens, GR-15771 Athens, Greece}
\author{C. Petri}
\email{cpetri@physnet.uni-hamburg.de}
\affiliation{Zentrum f\"ur Optische Quantentechnologien, Universit\"at Hamburg, Luruper Chaussee 149, 22761
Hamburg, Germany}
\author{P. Schmelcher}
\email{pschmelc@physnet.uni-hamburg.de}
\affiliation{Zentrum f\"ur Optische Quantentechnologien, Universit\"at Hamburg, Luruper Chaussee 149, 22761
Hamburg, Germany}
\date{\today}

\title{Critical non-equilibrium steady states of the Lorentz channel}

\begin{abstract}
We investigate the transport properties of non-interacting particles propagating in a finite Lorentz channel
(LC). We show that interparticle power-law correlations emerge, when the dynamics
is described at a spatially coarse-grained level. This behaviour
appears in the non-equilibrium steady state of the LC under flux boundary conditions and
persists even in the presence of external driving, provided that the billiard's horizon is infinite in a
static or temporal sense.  We show that Fermi acceleration permits the synchronization of particle motion with the
periodic appearance of the ballistic corridors, which, in turn,
gives rise to intermittent dynamics and the emergence of critical correlations. Thus, for the driven setup,
the critical state acts as an attractor
possessing characteristics of self-organization.
\end{abstract}
\pacs{05.60.Cd,05.45.Ac,05.45.Pq,05.65.+b,64.70.qj}
\maketitle

The Lorentz gas (LG) \cite{Lorentz:1905} acts in the theory of dynamical systems as a paradigm allowing us to
address fundamental issues of statistical mechanics, for instance, ergodicity and mixing
\cite{Sinai:1970,Bunimovich:1979,Bunimovich:1980}, as well as transport processes, such as diffusion in the
configuration space \cite{Machta:1983,Beijeren:1982,Gaspard:1998,Klages:2000,Armstead:2003}. The static
periodic LG comprises a regular lattice of circular fixed scatterers and an ensemble of non-interacting
particles traveling freely between collisions and scattering elastically off the circular obstacles. The
transport properties of such a system are determined by the billiard's geometry, that is the specific lattice
symmetry and the lattice constant. If the maximum free path length is not bounded from above, then the setup
possesses a so-called ``\textit{infinite horizon}'' (IH) and the diffusion in configuration space is anomalous
\cite{Bleher:1992,Szasz:2007}. For a more compact packing of the scatterers, i.e. ``\textit{finite horizon}''
(FH), arbitrarily long flights are not possible and the system exhibits normal diffusion \cite{Beijeren:1982}.
From a dynamical point of view, if the system has FH, then it is fully hyperbolic. However, in the case of the
IH geometry, the chaoticity weakens and an ordered portion of phase space emerges, due to the existence of
ballistic corridors, leading to an intermittent behaviour \cite{Artuso:2004}.
Independently of the specific geometry (FH or IH) it has also been rigorously proven that a finite LG under
flux boundary conditions possesses a nontrivial nonequilibrium steady state \cite{Gaspard:1997}.

Time-dependent generalizations of the original periodic Lorentz gas model have been introduced, in which
the scatterers are allowed to oscillate \cite{Loskutov,Karlis:2006,Oliveira:2011}, rendering the study of
diffusion in momentum space possible. This process is intimately linked to Fermi acceleration
\cite{Fermi:1949}, which is considered a fundamental acceleration mechanism in many areas of physics
\cite{Blandford:1987}. The mechanism consists in the increase of the mean energy of particles as a result of
random collisions with moving scatterers. 

In this Letter, we show for the first time the emergence of power-law (critical) correlations between the
propagating particles in the LG with IH in a channel geometry, despite the fact that they are non-interacting.
To reveal these cross-correlations we introduce a spatially coarse-grained description of the dynamics. This
critical behaviour appears in the nonequilibrium steady state established in the above setup, under flux
boundary conditions. We demonstrate the existence of such a nonequilibrium steady state in particular in the
presence of external driving and thereby, when the energy-shell is violated. We introduce the
dynamically infinite horizon (DIH) as a property of driven extended billiards for which ballistic corridors
-- choosing appropriate geometrical and driving parameters -- open up and close periodically in time,
i.e. exist only for certain time intervals. In this case, we show that the development of Fermi acceleration enables
the particles to synchronize their motion with the periodic appearance of the ballistic corridors, such that
they can perform free flights of arbitrary length, which, in turn, gives rise to intermittent dynamics and the
appearance of critical correlations. In this sense, it is shown that Fermi acceleration can act as an
effective driving force to steer an ensemble of propagating particles towards a nonequilibrium steady state possessing several
characteristics of self-organized criticality (SOC). This work adds to the recent intensive efforts to clarify
the interrelation between intermittency, criticality and self-organization
\cite{Antar:2001,Sattin:2006,Uritsky:2007,Sun:2010}.

\begin{figure}[htbp]
\includegraphics[width=8.6 cm]{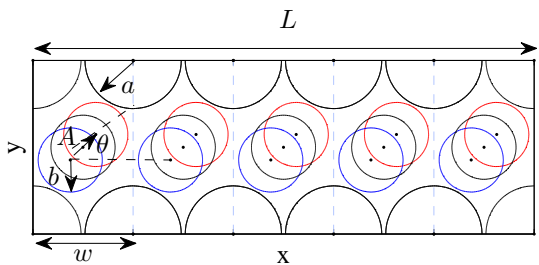}
\caption{\label{fig1} (color online). Schematic of the billiard geometry. The central scatterers of radius $b$
are driven
harmonically, while the scatterers with radius $a$ remain fixed. $A$ and $\theta$ are the amplitude
and angle of oscillation, respectively. The oscillating disks are depicted in their equilibrium
positions (black circles) together with the turning point positions (blue and red circles).} 
\end{figure}

The channel is constructed by superimposing two square lattices of circular scatterers, which have the same
lattice spacing $w$, and radii $a$ and $b$, respectively. The lattices are positioned such that there is a $b$
disk at the center of each unit cell of the lattice of $a$ disks. The system is finite in both the $x$ and $y$
direction.  In the $x$ direction it extents from $0$ to $L$, while two hard flat boundaries are placed at $\pm
w/2$, confining the motion of the particles in the $y$ direction. Particles are injected from the left end
with a uniform distribution of direction angles in $(-\pi/2,\pi/2)$ and a steady flow of $50$ particles per
unit time. When reaching either end of the channel the particles escape the system. Hence, the
considered device is a finite strip of a periodic LG, known as \textit{Lorentz channel} (LC)
\cite{Gaspard:1993,Sanders:2005,Alonso:1999}. Finally it is noted that the $a$ disks are fixed in space,
while the centers of the $b$ disks can oscillate harmonically with amplitude $A$, as shown in
Fig.~\ref{fig1}. In the following, length is measured in units of $w$ and time in units of $1/\omega$,
$\omega$ being the angular frequency of the oscillation of the central disks. Without loss of generality,
provided that the flat segments between the static discs are small enough to prohibit bouncing orbits in the
$y$ direction, we choose $a=0.48$ \footnote[1]{}. For the sake of brevity, in the following we analyze results
obtained in the steady-state of the driven LC, i.e. $A\ne0$.  We remark that our findings with respect 
to the stationary properties of the steady state of the driven LC hold in a similar way for the
static LC with IH \cite{Karlis:2011}. 

A derivation of the transport properties via the microscopic dynamics in such a system, even in the absence of driving, is impeded by the inherent
hyperbolicity. However, statistical aspects of the transport properties of the infinite
triangular static Lorentz gas, have been modeled by a suitable hopping process, either as a two-dimensional
completely uncorrelated random walk \cite{Machta:1983} or by taking into account correlations between jumps
\cite{Klages:2000,Klages:2002,Gilbert:2009}. Extending this description to the driven LG, and
bearing in mind that the billiard under
study is quasi-one-dimensional, $L\gg 1$, the transport properties can be studied by viewing the motion of the
particles as a hopping process between neighboring cells [marked by dashed blue vertical lines in Fig.~\ref{fig1}],
i.e. as an one-dimensional stochastic process. This coarse-graining of the billiard, leads to a
discrete chain consisting of $N=L$ sites. At this level, the motion of a particle inside a unit
cell -- the intra-cell dynamics -- is integrated out. Therefore, we concentrate on the
inter-cell properties of the motion of the particles.

Let us pose the question for the probability $P(n)$ that a particle 
will travel $n$ cells before changing its direction of the inter-cell motion. This probability  is
closely linked to the memory of the associated random walk process. To illustrate this, consider the time-independent LC,
where the central disks are fixed at the equilibrium positions of the time-dependent setup. If the horizon is
finite, i.e. $\sqrt{3}/2-a\le b$, then the system exhibits normal diffusion which can be quite accurately
approximated by a simple random walk, with jump probability $1/2$. In this case, $P(n)$ decays exponentially,
i.e. $P(n)\varpropto2^{-n}$ . On the contrary, if ballistic corridors
open up, the probability of long free flights between collisions is greatly enhanced and an algebraic decay of
$P(n)$ is expected \cite{Bouchaud:1985,Larralde:1998}, reflecting the long-range correlations in the
inter-cell dynamics.

\begin{figure}[htbp]
\includegraphics[width=8.6 cm]{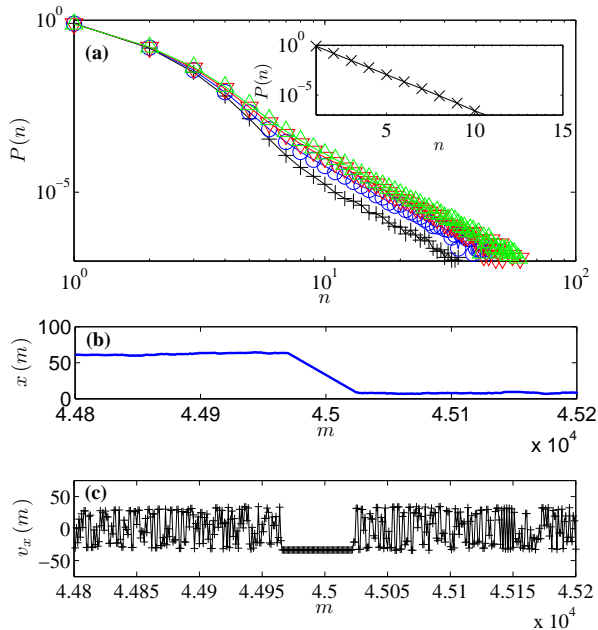}
\caption{\label{fig2} (color online). (a) The probability of crossing $n$ cells without changing direction of
motion with
$b=\sqrt{3}/2-a$, $\omega=1$ and $|\mathbf{A}|=1/2-b$ and $\theta=0.11$
[upright black crosses], $\theta=0.22$ [blue open circles], $\theta=0.33$ [red downright triangles]
$\theta=0.5$ [green upright triangles] and in the inset $P(n)$ for $\theta=0$, in units of $\pi$. (b) $x$
coordinate; (c) $v_x$ velocity of a typical trajectory versus the sum $m$ of the number of collisions with the billiard's
boundaries and crossings from cell to cell.}
\end{figure}

Let us now focus on the driven LC. In Fig.~\ref{fig2}(a) we present numerical
results for $P(n)$ obtained by the simulation of a LC
with 100 cells with a representative incoming flux \footnote[2]{}. The radii of the central disks are
$b=\sqrt{3}/2-a$, which is the
minimal value for the time-independent LC to possess a FH. The parameters of oscillation of the
moving disks are chosen as follows: The amplitude is fixed to $A=1/2-b$, which is the maximum allowed value
such that
the disk motion is contained within the cell boundaries, $\omega=1$, and we choose 5 different
values for $\theta$.
Evidently, $P(n)$ shows a power-law tail for $\theta\ne0$, which is the condition for the existence of
dynamically open ballistic corridors. However, for small $n$, $P(n)$ is exponential, due to the dominance of diffusive
transport. The value of $n$ for which the crossover from exponential to power-law behaviour takes place
depends on $\theta$, while the slope of the power-law tail is $\theta$-independent. For $\theta=0$ only an
exponential decay is observed, since ballistic
corridors remain closed for all times. Thus, for $\theta\ne0$ the particles synchronize
with the oscillation
of the central disks, such that long collisionless flights become possible. 
A detailed trajectory analysis verifies this and moreover, reveals that particles traveling long distances
without changing their direction of motion from cell to cell demonstrate an intermittent behaviour. This is
illustrated in Fig.~\ref{fig2}(b)-(c), where the $x$ coordinate of a particle and
the corresponding velocity component are shown as a function of the coarse-grained time variable $m$, taking
integer values. In fact $m$ is increased by $1$
either upon collision with the 
billiard's boundaries, or upon crossings across the border between neighboring cells. As observed, a
trajectory exhibits phases of chaotic motion, where it stays localized, interrupted by regular (ballistic)
motion. From a dynamical point of view, this corresponds to a transition from areas close to the ordered
portions of phase space to the hyperbolic areas owing to
the convex geometry of the billiard's boundaries. The regular part of the phase space exists in the high
energy regime for certain combinations of the phase of
oscillation of the moving disks and the direction of the velocities of the particles.

An interesting analogy can be drawn between the driven LC and the well-known sandpile model \cite{Bak:1987}
(for a review see Ref.~\cite{Turcotte:1999}). The redistribution of grains in the sandpile model, when out of
equilibrium, can be described by a diffusion equation \cite{Carlson:1990,Kadanoff:1991}. However, when
the sandpile reaches equilibrium the diffusion coefficient diverges and grains are redistributed with
avalanches, the probability of the number of sites participating in such events being a power-law. In the
driven LC
the inter-cell motion of the particles can be described by a random walk, with finite
moments of the
distribution of step lengths, when the particle motion is not synchronized with the moving scatterers. As a
result,
long ballistic flights are not possible and a diffusion coefficient can be defined. As the non-equilibrium
steady-state is reached, a subset of the ensemble of the particles propagating through the device will
approach the areas of the phase space corresponding to ballistic motion. From that point on, particle dynamics
will become intermittent and the simple random walk picture breaks down. The long flights
without collisions ---known as L\`{e}vy flights in the framework of random walks--- lead to the divergence of
the diffusion coefficient, as it is the case in the time-independent LC with infinite horizon
\cite{Larralde:1998}. A manifestation of the L\`{e}vy type statistics is the appearance of the power-law tail
of $P(n)$, similar to the appearance of ``avalanches'' in the sandpile SOC model. 
For the driven LC criticality implies the existence of ballistic corridors
at least temporally which depends on the geometrical details of the set-up and the
characteristics of the time-law of the driving. If ballistic corridors exist, then the
critical state is an attractor, without the need of a fine-tuning of the external parameters.

To be able to study the robustness of the critical state under the variation of system parameters, we
construct a complexity measure $C$, which is the product of two terms \cite{Lopez:1995}: the first term is the
quadratic distance of $P(n)$ from the exponential behaviour expected for a fully hyperbolic system; the second
term quantifies the amount of information ---in the Shannon sense--- stored in the system, not necessarily
related to complex behaviour. To obtain a normalized complexity measure $\tilde{C}$, $C$ is divided with its
maximum value obtained in the static, i.e. non-driven LC. By construction $\tilde{C}$ is positive definite,
maximizes when the probability of the occurrence of long collisionless flights becomes maximum and tends to
zero when the motion of particles in the systems is completely chaotic [exponential $P(n)$].
\begin{eqnarray}
\label{eq1}
 C&=&-\sum\limits_{n=1}^N\left[\log\left(P(n)\right)-\log\left(\frac{2^{-n}}{
1-2^{-N}}\right)\right]^2\times\nonumber\\
\label{eq1b} &&\sum\limits_{n=1}^N\frac{P(n)\log(P(n))}{\log(N)}.
\end{eqnarray}

\begin{figure}[htbp]
\includegraphics[width=8.6 cm]{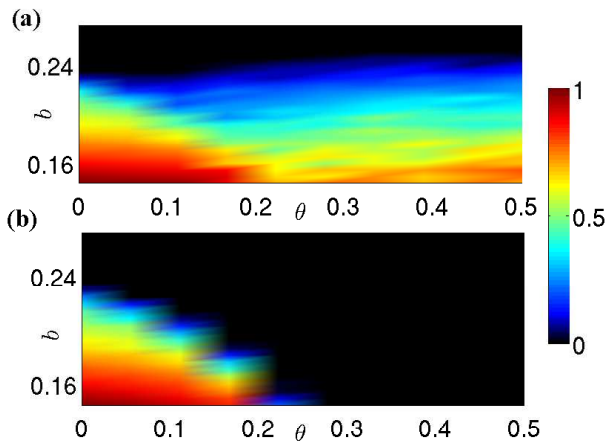}
\caption{\label{fig3} (color online). A contour plot of $\tilde{C}$ constructed by the
simulation of the billiard's dynamics for 25 different values of $b$ and 10 values of $\theta$. 
The black area
corresponds to the non-critical state (exponential $P(n)$). Panel (a) corresponds to synchronous oscillation
of the central disks while (b) implies uniformly distributed random initial phases. Note that in both cases,
as in Fig.~\ref{fig2}, $|\mathbf{A}|=1/2-b$, which is the maximum allowed value of the oscillation
amplitude $\vert\mathbf{A}\vert$ for a given $b$.} 
\end{figure}

In Fig.~\ref{fig3}, we present contour plots of the measure $\tilde{C}$ as a function of the radii of the
oscillating disks $b$ and the angle of oscillation $\theta$. In Fig.~\ref{fig3}(a), the oscillation of the
central disks is synchronous, whereas in Fig.~\ref{fig3}(b) the initial phases of the oscillating disks are
chosen randomly according to a uniform distribution in the interval $[0,2\pi)$. As seen in
Fig.~\ref{fig3}(a),
the system always reaches the critical state, unless the parameters of oscillation and the radius $b$ are
such that ballistic corridors remain closed for all times. On the contrary, when the initial phases are
randomly chosen dynamically open ballistic corridors cannot appear. Therefore,
ballistic corridors exist only in the static sense, i.e. corridors remain continuously open despite the motion
of the central disks. Thus, from geometrical considerations, the ``phase boundary'' is
$\sqrt{3}-2a-(1-2b)\sin\theta-2b=0$.

\begin{figure}[htbp]
\includegraphics[width=8.6 cm]{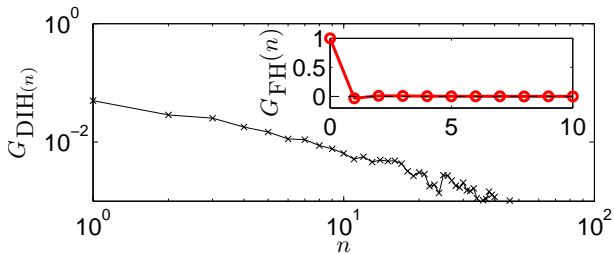}
\caption{\label{fig4} (color online). The cross-correlation function $G_{\text{DIH}}$ between independent 
trajectories in an
DIH setup, with $b=0.16$. In the inset $G_{\text{FH}}(n)$ is also plotted for a set-up with
FH $b=a=0.48$.}
\end{figure}

Given that the particles propagating through the LC are non-interacting, the question arises as to how the
system can reach a critical state. The scale-free behaviour, as demonstrated by the power-law decay of
$P(n)$ at criticality, emerges at the level of the statistical ensemble because of the intermittency owing to
the dynamically open ballistic corridors. These ordered  structures deform the geometry of the phase space
in such a way that correlations between independent trajectories, starting from different
initial conditions, are induced. To verify this, we performed a symbolic dynamics analysis using a binary
alphabet $\mp1$, corresponding to a change or not of the direction of motion each time a particle crosses a
cell. In Fig.~\ref{fig4} we present the correlation function computed on the basis of a long sequence,
containing $\sim5\times10^6$ symbols, constructed by tracking a large number of trajectories. We observe that
when the driven LC has a DIH (critical state) the correlation function $G_{DIH}(n)$ decays algebraically.
On the contrary, when the system is away from criticality (FH), the correlation function $G_{FH}$ drops
by several orders of magnitude in just a single step.

In conclusion, we have shown that power-law (critical) correlations emerge between non-interacting
particles propagating in the LC. This behaviour, revealed at a suitable spatial scale, is established in the
nonequilibrium steady state reached in the static and driven LC. The key property leading to criticality is
intermittency, which, combined with ergodicity,
generates power-law cross-correlations.  Intermittency implies the mixing of ordered with chaotic parts of
trajectories at all time-scales,
while ergodicity is responsible for transferring trajectory properties to the statistical ensemble. The
interrelation of intermittent dynamics and critical behaviour has already been pointed out in
Ref.~\cite{Contoyiannis:2002}, in the framework of the 3D Ising model. However, herein the mechanism leading
to power-law correlations is reduced to its basic ingredients, due to the absence of any inter-particle
interaction in the considered ensemble. This goes one step beyond the model of self-driven particles
\cite{Vicsek:1995,Huepe:2004}, where inter-particle interactions are substituted by certain rules, leading to
correlations. In this sense, this work sheds new light on the emergence of power-law correlations and
criticality in dynamical systems in general. Finally, in the driven LC with DIH, we demonstrate a novel
role of Fermi acceleration as
an effective driving force steering an ensemble of propagating particles towards the critical state inducing
this way characteristics of self-organization.

\begin{acknowledgments}
This work was made possible by the facilities of the Shared Hierarchical Academic Research Computing Network
(SHARCNET:www.sharcnet.ca) and Compute/Calcul Canada. This research has been co-financed by the European Union
(European Social Fund --- ESF) and Greek national funds through the Operational Program "Education and
Lifelong Learning" of the National Strategic Reference Framework (NSRF) - Research Funding Program:
Heracleitus II. Investing in knowledge society through the European Social Fund. The authors also thank the IKY and DAAD for financial support in the framework of an exchange program between Greece and Germany (IKYDA).
\end{acknowledgments}

\end{document}